\documentclass[aps,preprint,superscriptaddress,showpacs]{revtex4}
\usepackage[english]{babel}
\usepackage{braket}
\usepackage{slashbox}
\usepackage{epsfig}
\usepackage{amsfonts}
\usepackage{graphicx}
\usepackage{hyperref}
\usepackage{natbib}
\usepackage{amsthm}
\usepackage{amsmath}
\usepackage{color}
\usepackage{times}
\usepackage{psfrag}
\usepackage{subfigure}
\usepackage{epstopdf}
\begin{document}

\title{Coherence and stochastic resonances in a noisy van der Pol-type circadian pacemaker model driven by light}

\author{\textbf{F. L. Tsafack Tayong}}
\affiliation{Fundamental Physics Laboratory, Physics of Complex System group,
Department of Physics, Faculty of
 Science, University of Douala, Box 24 157 Douala, Cameroon.}
\author{\textbf{R. Yamapi}}
\email[ryamapi@yahoo.fr]{(Corresponding author)}
\affiliation{Fundamental Physics Laboratory, Physics of Complex System group,
Department of Physics, Faculty of
 Science, University of Douala, Box 24 157 Douala, Cameroon.}
\author{\textbf{G. Filatrella}}
 \affiliation{Department  of Sciences and Technologies
\small and INFN Gruppo collegato Salerno, University of Sannio, Via de Sanctis,
I-82100 Benevento, Italy.}
\date{\today}

\begin{abstract}
Daylight plays a major role in the wake/sleep cycle in humans.
Indeed, the wake/sleep system stems from biological
 systems that follow a circadian rhythm determined by the light/dark alternation.
The oscillations can be modeled by the higher order non-linearity van der Pol -type
equation  driven by a term that mimics the light cycle.
In this work noise in the illumination is introduced to investigate its effect on the human circadian cycle.
It is found that the presence of noise is detrimental for the
sleep/wake rhythm, except for some special values for which it may favor regular oscillations.
Depending for system parameters,  noise  induces regularities,
such as stochastic resonance: if the
natural light is turned off, it emerges that there is an
optimal value of intensity noise which most deteriorates the regularity
of the cycle, it is the phenomenon of anti-coherent resonance.
Also, the phenomenon of stochastic resonance occurs: in the
presence of the drive of natural light, there is an optimal
noise intensity which improves the evolution of the wake / sleep system.
However, there is a critical value of the noise beyond which
the system becomes chaotic; indeed, for sufficiently high noise
levels (how high depends upon the parameter of the system),
the sleep/wake cycle evolves in a random and unpredictable manner,
for whatever parameters of the external light.

\end{abstract}
\maketitle


\section{Introduction}
\label{Introduction}

A biological rhythm is defined as a series of statistically
significant physiological variations that appear as time oscillations of reproducible form.
Circadian rhythms, characterized by a period close to 24 hours, 
are observed in almost all living organisms, bacteria, mammals, 
plants and insects \cite{Lit1, Lit2}. Of all the organisms studied,
 the human organism in particular is synchronized thanks to an internal 
 clock located at the level of the brain, in the supra-chiasmatic 
 nuclei of the hypothalamus \cite{Lit33}; these nuclei are organs 
 located on either side of the third ventricle just above the chiasma, as their name suggests.

Deterministic models based on self-sustaining van der Pol
oscillators such as the Kronauer \cite{Kronauer99b} and
Jewett \cite{Jewett98} models account for the existence of
 circadian rhythms under constant environmental conditions.
The Jewett's variation \cite{Jewett98} of the higher order nonlinearity
 van der Pol oscillator  that is highly tractable also in 
 presence of noise \cite{Yamapi10,Yang18} and periodic 
 drive \cite{Mbakob17} that  is added to mimic the external
  light \cite{tsaf}, as the changes in light have an influence on the internal circadian clock.
Thus, light gives the circadian clock an indication about the time of the day.

In the biological system, noise is naturally present and 
essentially unavoidable, for biological signals always act on a background of disturbances.
Noise generally describing the undesirable fluctuations 
and being able to change the dynamics of a system, it is 
thus a principal question to study the chaotic behavior consequent to the presence of noise.
The aim of the paper is to determine the conditions of 
stability in the system and subsequently to know if noise 
can have a constructive influence on the evolution of the 
studied system (as e.g. in the noise induced enhanced 
stability \cite{Spagnolo15}); in other words, can noise 
contribute to the correct process of the sleep-wake cycle 
in living beings? Of course, noise in the illumination is a phenomenon
 to living organisms, for it has appeared with the appearance of artificial light.

The name of a phenomenon which is a prime example of the 
counter-intuitive idea of non monotonic consequences of a 
random term is Stochastic Resonance (SR) \cite{Gammaitoni98,Lit3}. 
Thus, we shall emphasize  the phenomenon of resonance 
(coherent or stochastic) on the wake-to-sleep cycle.
Stochastic Resonance, although a term originally used 
in the specific context of noisy and driven bistable systems, 
is now widely applied to describe any phenomenon where the right
 amount of noise in a nonlinear system improves the quality of
  the response \cite{Lit4}. Stochastic resonance has been widely
   observed in nature; it has been reported and quantified in systems 
   as diverse as electronic circuits \cite{Lit5}, neural models \cite{Lit6}, 
   chemical reactions \cite{Lit7}, cell biology \cite{Lit8} and even organic 
   chemistry semiconductors \cite{Lit9} and superconducting elements \cite{Addesso16}.

The paper is organized as follows: in section \ref{model}, we briefly describe the biological model of the van der Pol oscillator, the added external force and the noise introduced into the system
 to highlight some digital tools useful for our study; in Section \ref{Dynamics} we study the chaotic behavior of the system, in presence of random fluctuations, as a function of the parameters and of the external excitation; in Section \ref{coherence} we bring out the phenomenon of coherent resonance in the presence of noise;
next we focus on stochastic resonance in Section \ref{stochastic} for the study of resonance;
finally, Section \ref{conclusions} leads to the conclusions.

\section{The Van der Pol-type circadian model driven by light and noise}
\label{model}

 For several years, some mathematical models of the circadian stimulator of the human body have been made, among which we can refer to the two-process model \cite{model1} and the model based on van der Pol's oscillators \cite{Jewett98}.
 In this Section, it will be given a complete description of the model used, a variation of the van der Pol type oscillators  to account for some realistic features of the light drive, and  some tools necessary for its numerical solution will be highlight.

\subsection{The van der Pol-type circadian model driven by light}

The wake/sleep cycle model used is the basic model van der Pol \cite{Kronauer82, Kronauer90, Jewett99} self-sustaining oscillator with order seven nonlinearity described by the following differential equation:
\begin{equation}
\label{eq1}
  \ddot{x}+\mu\left(-\frac{1}{3}-4x^{2}+\frac{256}{15}x^{6} \right)\left(\frac{\pi}{12}\right)\dot{x} +\left(\frac{2\pi}{\tau_{x}} \right)^{2}x = 0,
\end{equation}
 where $ x $ represents the amplitude of the circadian stimulator which is closely linked to all the mechanisms of the human body (internal body temperature, the desire to sleep, level of vigilance, and so forth) having a daily period \cite{Kronauer82}, $ \tau_{x} $ the period of the pacemaker which is approximately $ 24.4h $ and $ \mu $  the stiffness of the oscillator chose in the interval $ ] 0; 0.45] $.
Since the pulses of light applied at the right time in a circadian cycle have a direct effect on the internal clock, one can add to Eq. (\ref{eq1}) the luminous flux $ I (t) $ as:
\begin{equation}
\label{eq11}
  \ddot{x}+\mu\left(-\frac{1}{3}-4x^{2}+\frac{256}{15}x^{6} \right)\left(\frac{\pi}{12}\right)\dot{x} +\left(\frac{2\pi}{\tau_{x}} \right)^{2}x = I(t),
\end{equation}
we note that $ I (t) $ consisting of a mixture of an empty function representing total darkness during the night and a sinusoidal function representing the evolution of sunlight during the day \cite{tsaf}.
A model of this type of illumination $ I (t) $ is defined as follows:
\begin{eqnarray}
\label{forcing}
I(t) &= \theta\left(D_L - t_{24}\right) I_0 \sin{\left(\omega  t_{24} +\phi \right)}  \\
 \omega &= \frac{\pi}{D_L}, \quad \quad  t_{24} = {\textrm mod} (t,24) \nonumber ;
\label{newperiod}
\end{eqnarray}
here $ I_{0} $ is the amplitude of the light intensity,
$ 24h $ is the cycle period, $ D_L $ is the duration of
the daylight, $ \theta $ the Heaviside function and $ \phi $
 the difference of phase between light and circadian stimulator.
 The model so far introduced employs arbitrary units.
However, for the sake of intuition, time is normalized to $24$ rather than to $1$.

\subsection{The stochastic van der Pol-type circadian model}

Any set of unwanted elements may be considered noise to be added to a signal altering its characteristics.
Noise can be introduced in a multiplicative or additive way \cite{Nagumo}; moreover it can white or colored \cite{Zaikin}.
To make the system more realistic, a random excitation can be thus added to the model of the sleep-wake cycle of Eq. (\ref{eq11}), which amounts to the following equation:
\begin{equation}
\label{eq2}
  \ddot{x}+\left(\frac{2\pi}{\tau_{x}} \right)^{2}x +\mu\left(-\frac{1}{3}-4x^{2}+\frac{256}{15}x^{6} \right)\left(\frac{\pi}{12}\right)\dot{x} =  \theta\left(  I(t) + \eta (t) \right)\left[ I(t) + \eta (t) \right].
\end{equation}
$ \eta (t) $ here characterizes the additive Gaussian white noise \cite{blanc, blanc1} which is an elementary noise model used to mimic many random processes that occur in nature. The Heaviside function is introduced to ensure that the total illumination is positive.

Noise follows the normal laws of mean and variance of data, with the following properties:

\begin{subequations} \label{eq3}
  \begin{align}
    \langle \eta (t)\rangle &= 0 ,  \label{eq3a}\\
    \langle \eta (t),\eta (t^{'})\rangle &= 2 D \delta(t- t^{'}),  \label{eq3b}
    \end{align}
\end{subequations}
where $ D $ governs the magnitude of the noisy external force, $\delta$ is the Dirac function.
Several natural sources produce white noise that can be assumed uncorrelated, one can cite thermal vibrations linked to atoms in conductors, bodies or objects at high temperature, celestial sources and the black body radiation of the Earth \cite{blanc2}.
We here assume that artificial light is of this type.

\subsection{Algorithm of numerical simulations}

Numerical simulations are based on the implementation of mathematical techniques in order to solve real and complex physical problems which are most of the time difficult to solve analytically.
Thus, to generate the white noise we are going to use the Box-Mueller algorithm \cite{fox} through two random numbers uniformly distributed over a unit interval, that gives the following algorithm:

\begin{subequations} \label{eq21}
  \begin{align}
a,b &= \mathrm{random  \, numbers,} \label{eq21a}\\
\eta_{h} &= \sqrt{ [-4 D h \log (a) ]} \cos(2\pi b) , \label{eq21b}  \\
 x_{t+h} &= x_{t} + h u_{t}  , \label{eq21c}\\
u_{t+h} &= u_{t} + h \left[-\mu\left(-\frac{1}{3}-4x_{t}^{2}+\frac{256}{15}x_{t}^{6} \right)\left(\frac{\pi}{12}\right)u_{t} -\left(\frac{2\pi}{\tau_{x}} \right)^{2}x_{t} \right]+ h\quad \theta\left(  I(t) + \frac{\eta_{h}}{h} \right) \left[ I(t)  + \frac{\eta_{h}}{h} \right] ,  \label{eq21d}
\end{align}
\end{subequations}
where $h$ is the integration time step, generally very small to ensure
the stability of the system ($ h = 10^{- 3} $ in the simulations presented here).


\section{Dynamical Analysis}
\label{Dynamics}

Dynamical bifurcation is concerned with a family of random dynamical systems which
is differential and has the invariant measure  $\mu_\alpha$.
If there exists a constant $\alpha_D$  satisfying in any neighbourhood of $\alpha_D$,
there exists another constant $\alpha$ and the corresponding
invariant measure  $\nu_\alpha\neq \mu_\alpha$ satisfying  $\nu_\alpha\to\mu_\alpha$ as
$\alpha\to\alpha_D$. Then, the constant $\alpha_D$  is a point of dynamical bifurcation.

\begin{figure}
\begin{center}
\includegraphics[height=6.0cm,width=14.0cm]{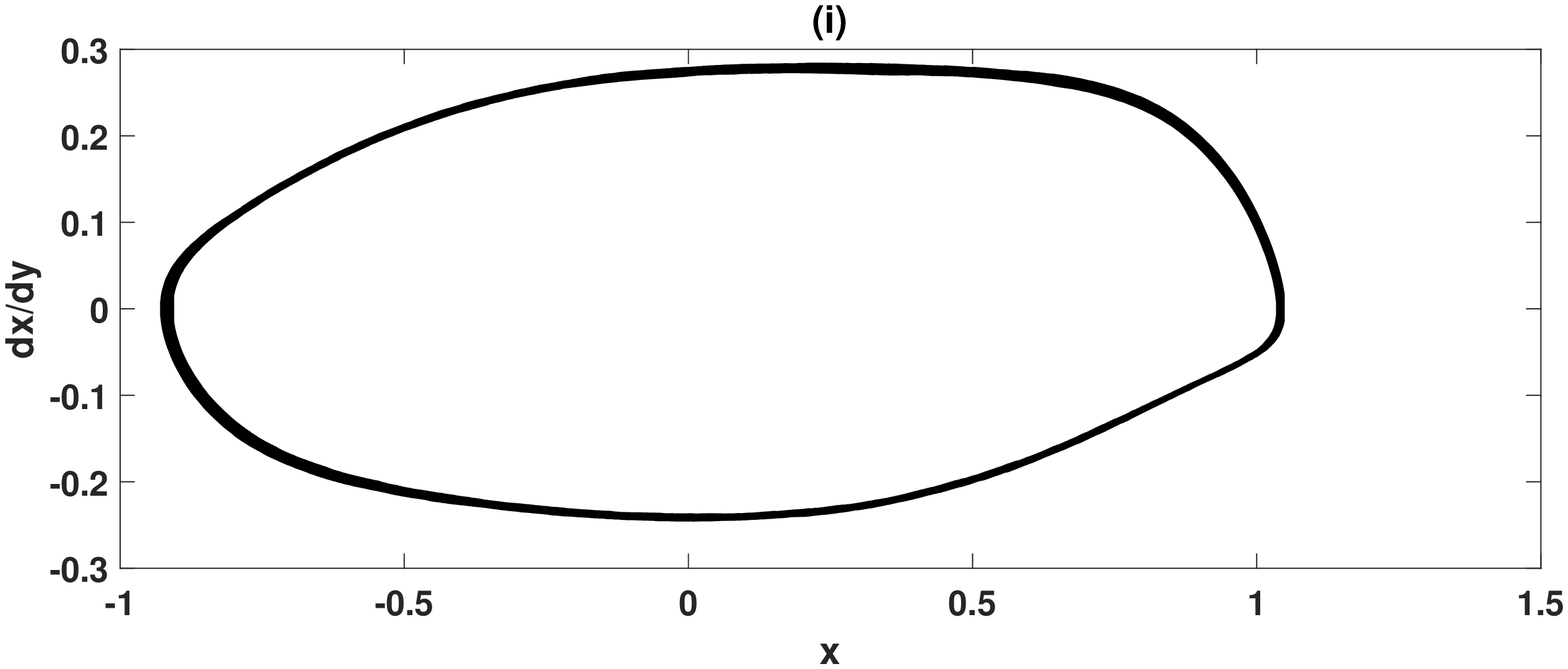}
\includegraphics[height=6.0cm,width=14.0cm]{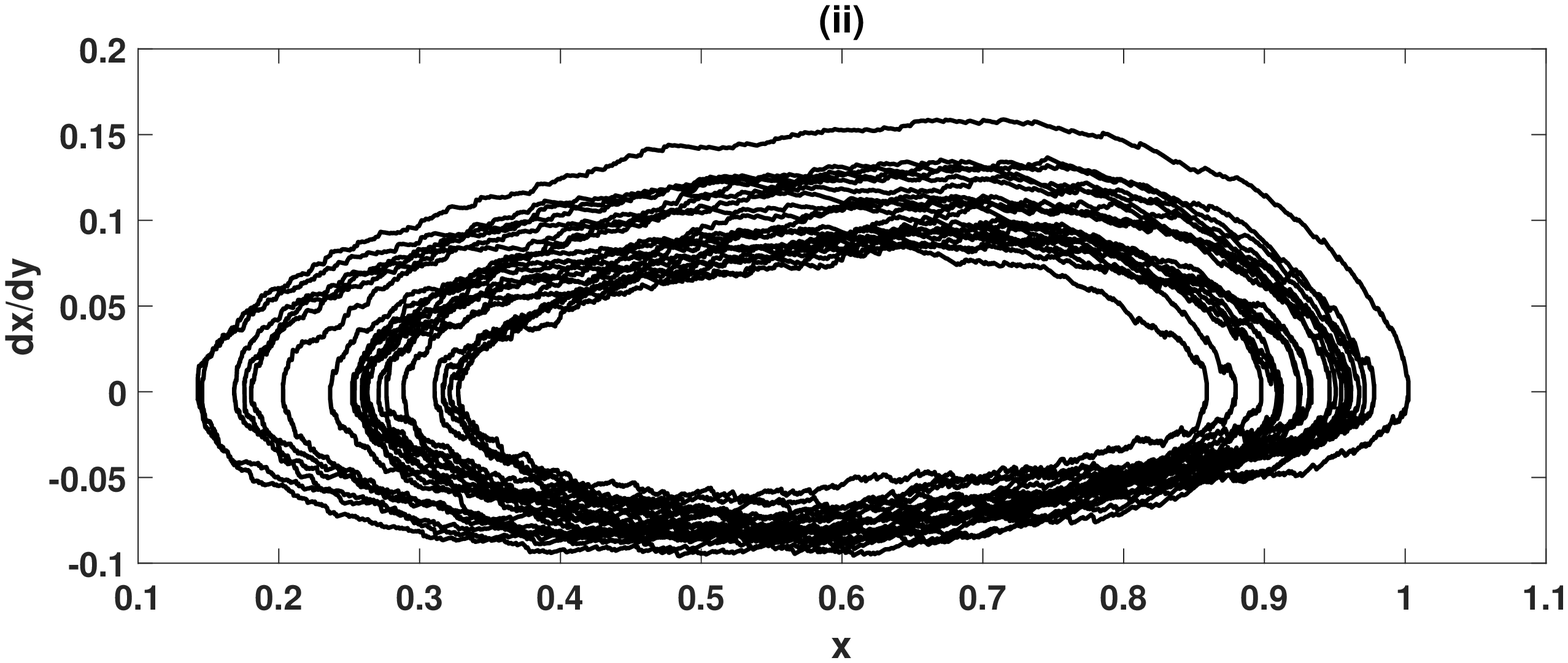}
\includegraphics[height=6.0cm,width=14.0cm]{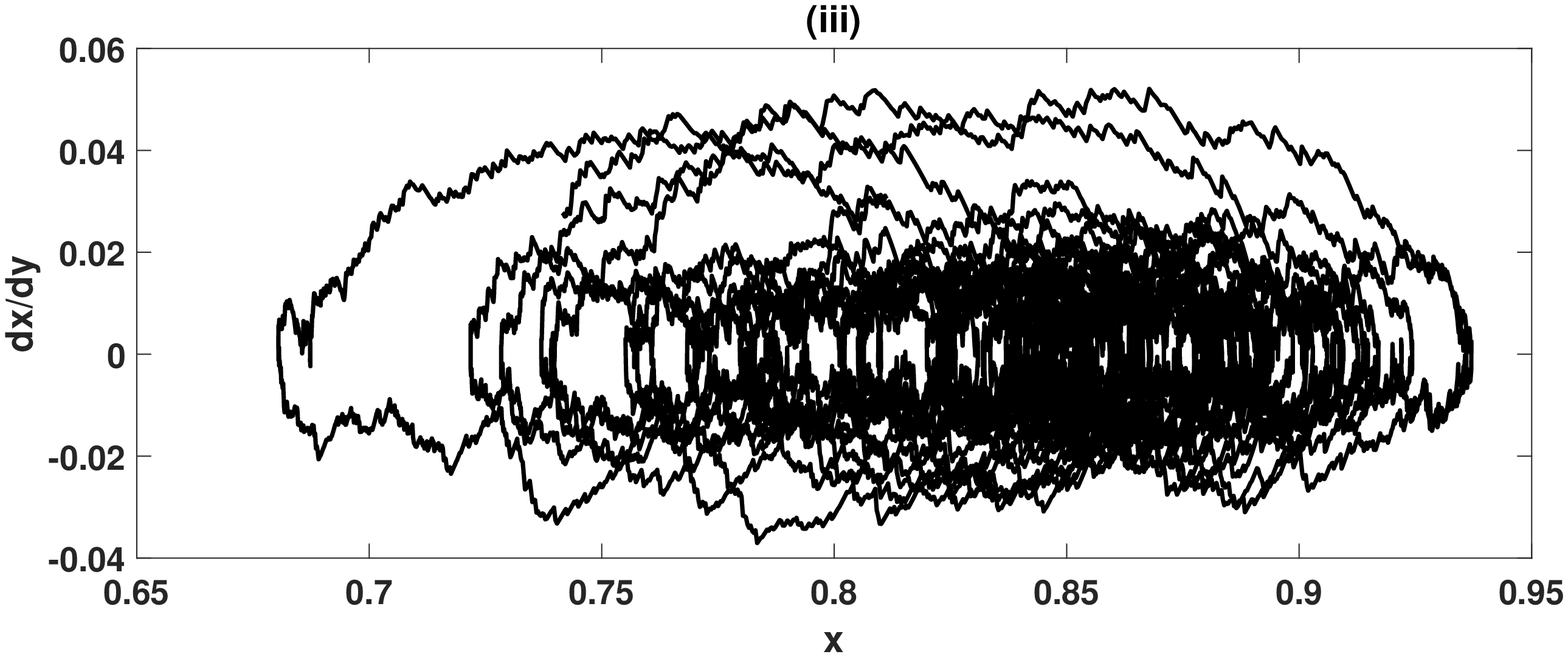}
\caption{\it  Phase portrait in a noisy van der pol-type circadian pacemaker driven by light
 for  $ D= 10^{-6} $ (i), $ D= 5  $ \textcolor{red}{$ \times $}$  10^{-5} $ (ii) and $ D= 10^{-4}$ (iii). ( $I_{0} = 0$, $ \mu = 0.23$ ) .
 The other parameter is: $\tau_{x} = 24.2 $.}
\label{fig1}
\end{center}
\end{figure}

\begin{figure}
\begin{center}
\includegraphics[height=6.0cm,width=14.0cm]{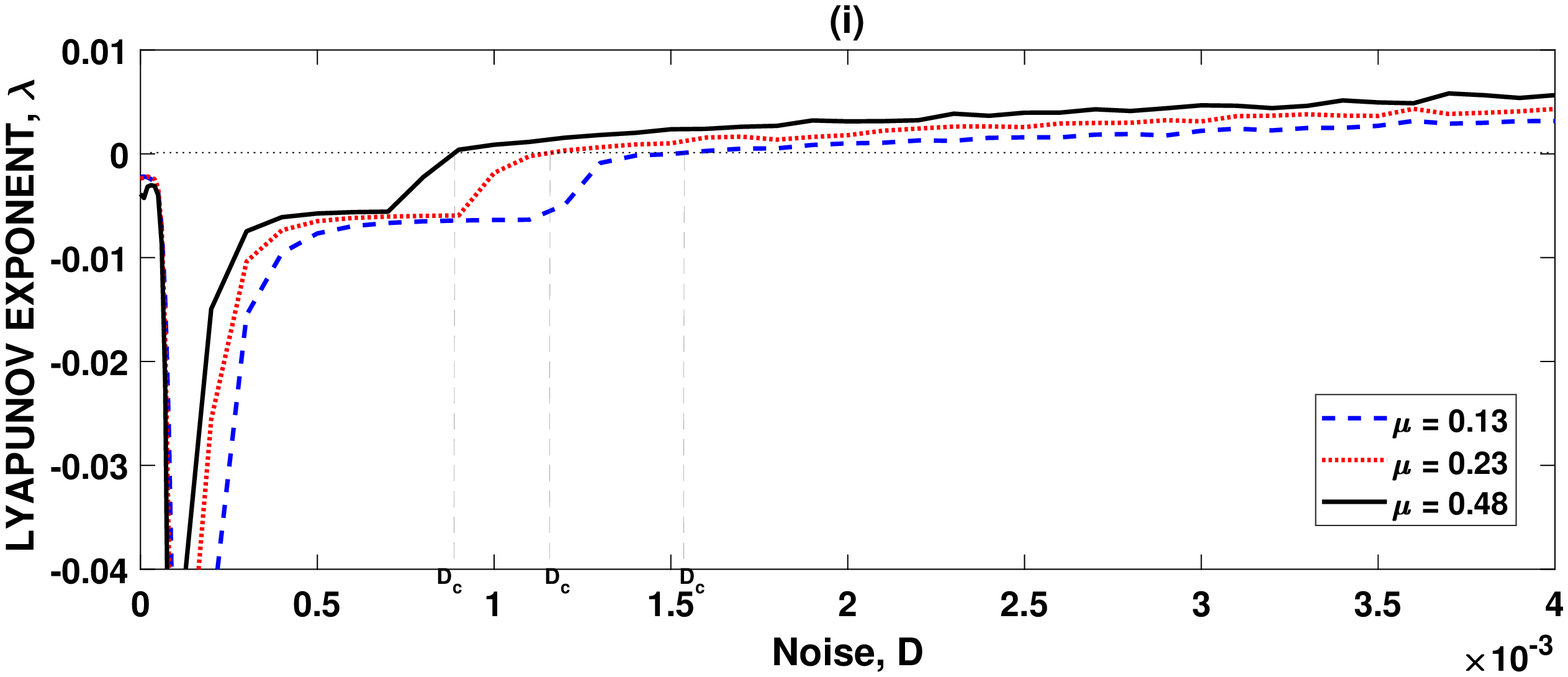}
\includegraphics[height=6.0cm,width=14.0cm]{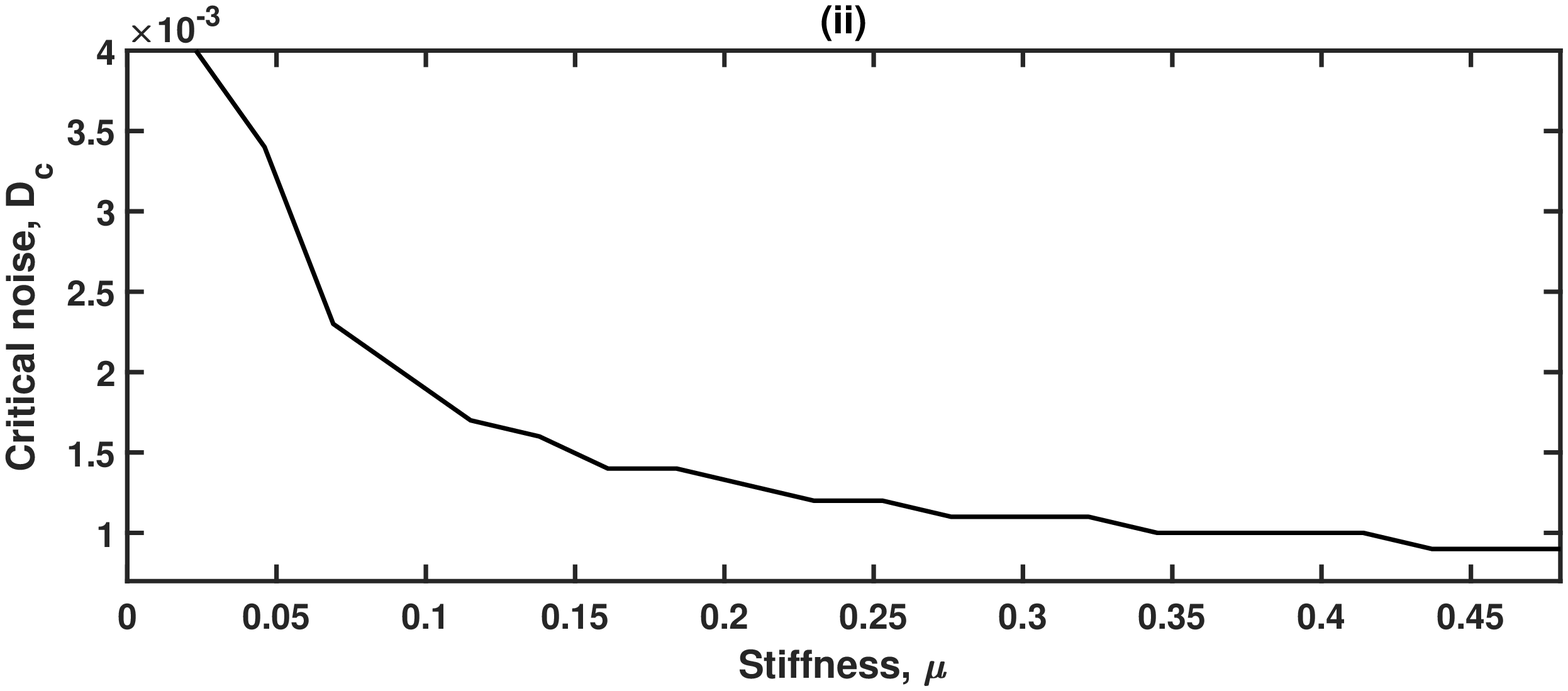}
\caption{\it  (i) Effects of the stiffness $\mu$ on the variation of the largest Lyapunov exponent
versus the intensity of noise $D$ in the absence of the light intensity  $ I_{0} = 0 $. (ii)
Critical noise boundary in the ($\mu $,$ D_{c}$) plane.
 The other parameter is: $\tau_{x} = 24.2 $.}
\label{fig2}
\end{center}
\end{figure}

In this section, one finds the effects of noise in
the higher nonlinearity van der pol-type circadian pacemaker model driven by light.
Dynamical bifurcation (called D-bifurcation) is analysing using
 the largest Lyapunov exponent $\lambda$ is defined as \cite{lyap1,lyap}:
 \begin{equation}
\label{eq11}
\lambda=\lim_{t\to
\infty}\frac{1}{t}\ln\left|\frac{\delta\mathbf{X}(t)}{\delta\mathbf{X}(0)}\right|=\lim_{t\to
\infty}\frac{1}{t}\ln\left|D\Phi\right(\mathbf{X})|,
\end{equation}
where $\Phi$ is the flow of the \textcolor{red}{ van der pol model } and $D\Phi(\mathbf{X})$, the Jacobian of $\Phi$ defined at the state $\mathbf{X}$.
The analysis of the stability of the system is carried out in the presence of additive Gaussian white noise, which is then related to the abrupt change of the sign of the  Lyapunov exponent.
Figure \ref{fig1} represents the phase portraits for some small noise intensity amplitude,
it is clear that the system becomes more and more disordered as noise increases; which makes perfect sense as noise is supposed to be harmful to physical ordered  systems.
We firstly investigate the influence of the noise intensity
$D$ on the small parameter $\mu$, seeking for sudden changes of
the maximum Lyapunov exponent, or D-Bifurcations.
Indeed, in Fig. \ref{fig2} (i) it is shown the variation
of the largest Lyapunov exponent versus the noise intensity
 for $ I_{0} = 0 $ for different values of the nonlinear damping $ \mu $ of the system.
In this figure one can notice that there is a value of the noise beyond which the system becomes chaotic.
Indeed, as the noise reaches the value $ D_{c} $, the Lyapunov number becomes positive and therefore the system becomes chaotic.
This critical value depends on the stiffness $ \mu $ of the system.
To be more detailed, we have drawn in Fig. \ref{fig2} (ii) the map ($ \mu $, $ D_{c} $) representing the critical noise as a function of the stiffness of the system. It turns out that when the value of $ \mu $ increases, the interval of noise amplitude $] 0; D_{c}] $ in which the system is stable decreases; thus the increase in stiffness contributes to the a loss of stability of the wake/sleep system in the presence of noise.

\begin{figure}
\begin{center}
\includegraphics[height=6.0cm,width=14.0cm]{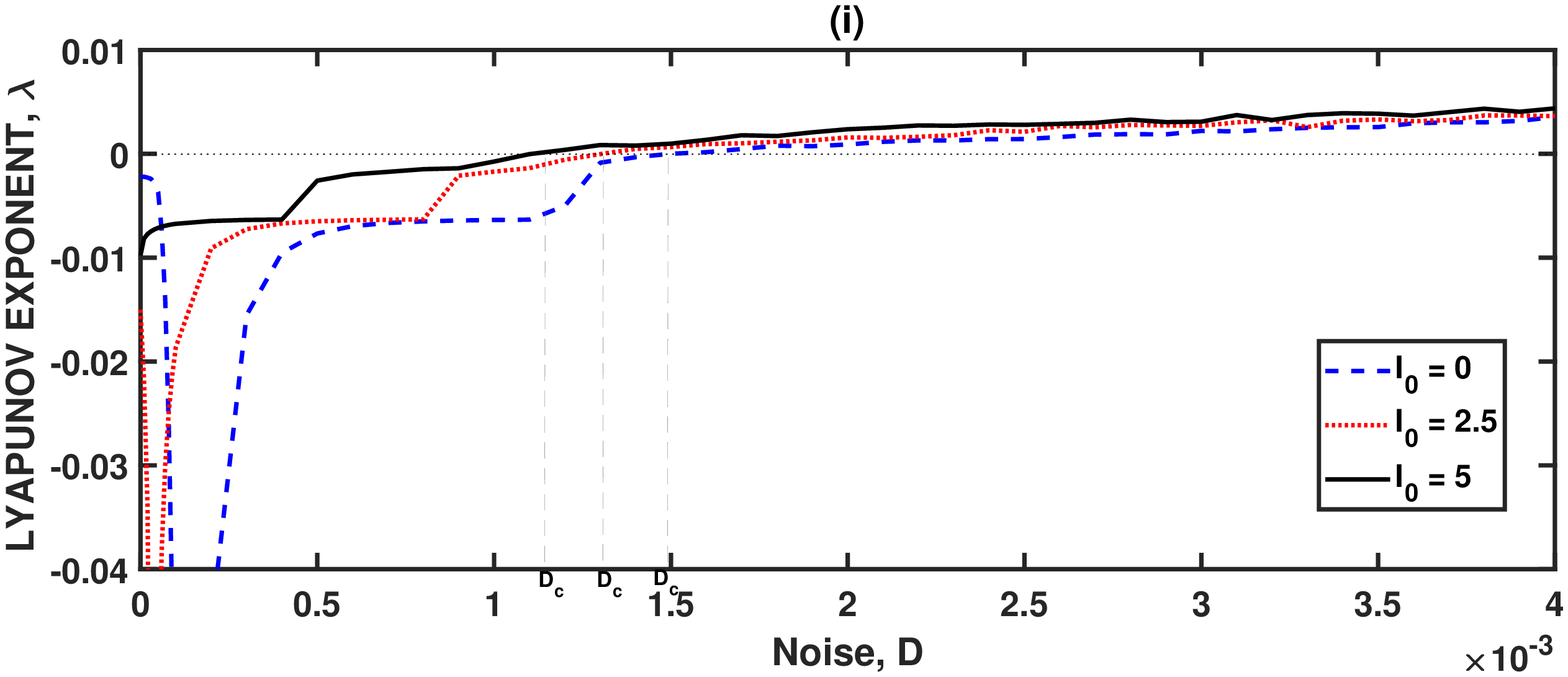}
\includegraphics[height=6.0cm,width=14.0cm]{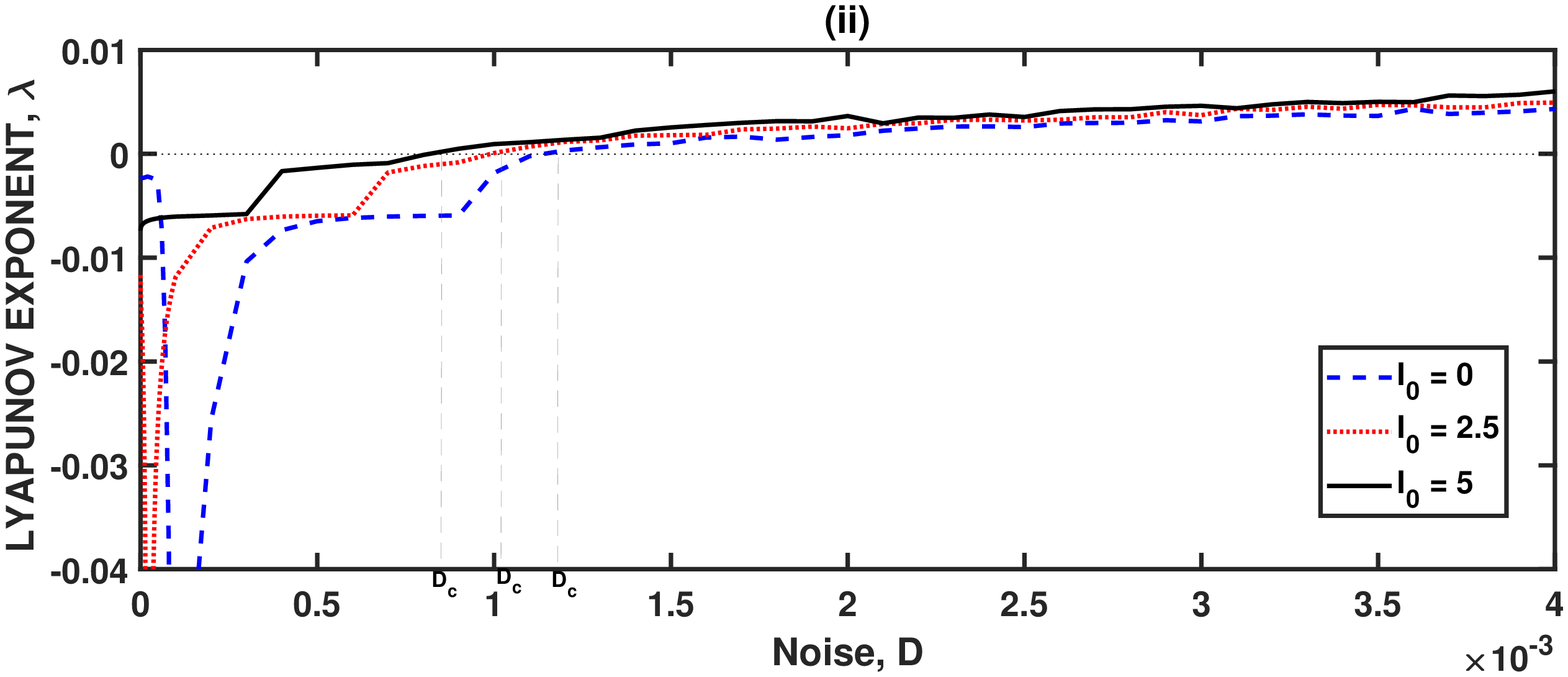}
\includegraphics[height=6.0cm,width=14.0cm]{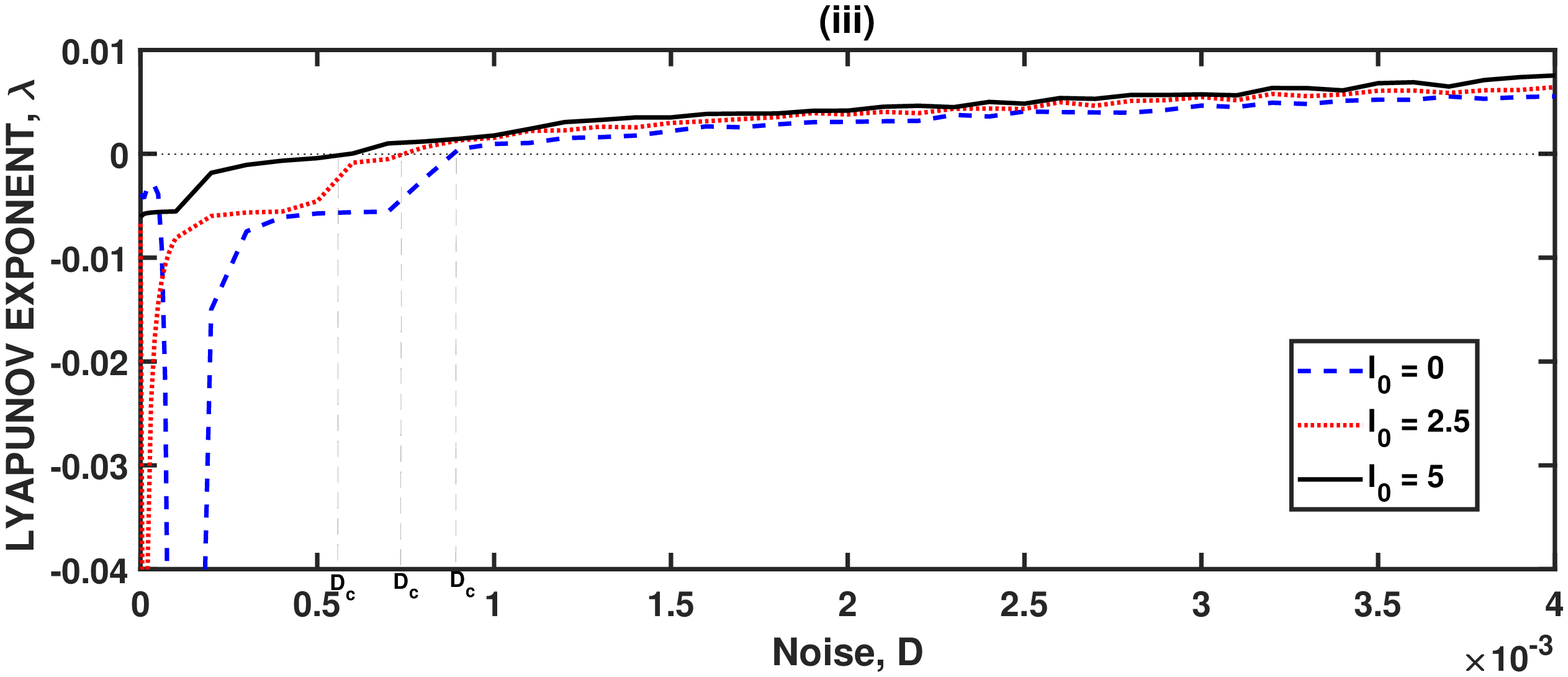}
\caption{\it Effects of the noise intensity $I_0$ on the variation
of the largest Lyapunov exponent versus the noise intensity  $D$ for several different
values of the stiffness coefficient;
(i): $\mu = 0.13$;$ (ii):\mu = 0.23$; and  (iii):$ \mu = 0.48$.
The other parameters are : $ D_{L} = 12 h $, $ \phi = 0 $, $\tau_{x} = 24.2$.}
\label{fig3}
\end{center}
\end{figure}

\begin{figure}
\begin{center}
\includegraphics[height=7.0cm,width=14.0cm]{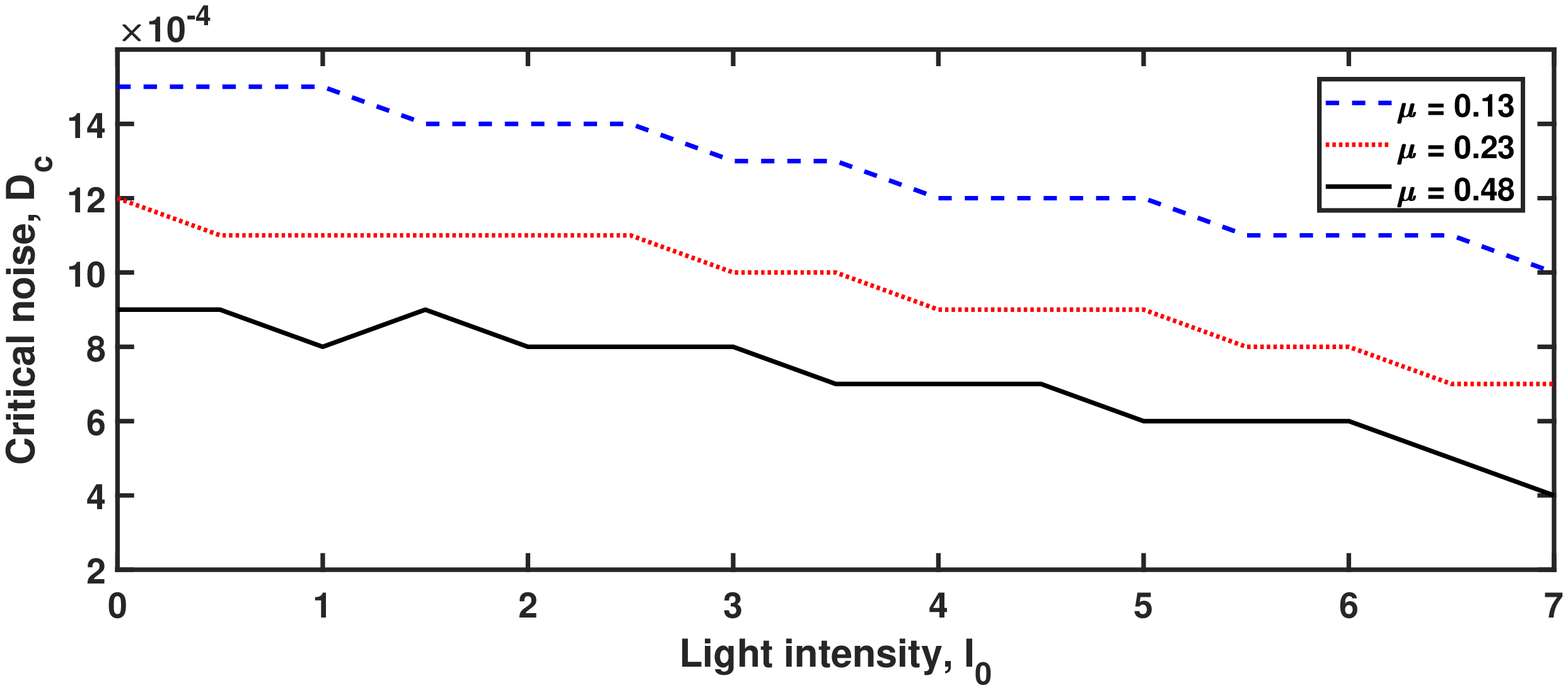}
\caption{\it Effects of the stiffness on the critical noise boundary in the
  $(I_{0},D_{c})$ plane.
The other parameters are : $ D_{L} = 12 h $  ,$ \phi = 0 $ and  $\tau_{x} = 24.2$ .}
\label{fig4}
\end{center}
\end{figure}

Figures \ref{fig3} show the variation of the largest Lyapunov exponent in the presence of the noise $ D $ and the light intensity $ I_{0} $ of the external force for three values of the stiffness ($ \mu = 0.13 $ i , $ \mu = 0.23 $ ii and $ \mu = 0.48 $ iii).
The light intensity has an impact on the critical noise $ D_{c} $; indeed the increase of $ I_{0} $ causes the decrease of $ D_{c} $.
This critical noise is a function of the stiffness $ \mu $ of the system.
In fact, we have drawn in figure \ref{fig4} the map ($ I_0; D_c $) of
the critical noise as a function of the light intensity for three values
of the stiffness. It is therefore clear that the light intensity and
the stiffness have an impact on the critical noise.

   \begin{table}[h!]
        \begin{center}
           \begin{tabular}[b]{|l|c|c|c|c|c|c|}
                        \hline
            \backslashbox{$I_{0} $ }{$ \mu$}& 0.023  & 0.069 &  0.115   & 0.23  & 0.39     & 0.46 \\
                         \hline
                      0.5 &  3.7 \textcolor{red}{$ \times $} $10^{-3}  $    & 2.29 \textcolor{red}{$ \times $} $10^{-3}  $ & 1.6 \textcolor{red}{$ \times $} $10^{-3}  $  &  1.19 \textcolor{red}{$ \times $} $10^{-3}  $  & 1.03 \textcolor{red}{$ \times $} $10^{-3}  $ &  0.89 \textcolor{red}{$ \times $} $10^{-3}  $   \\
                      \hline
                        1 & 3.69 \textcolor{red}{$ \times $} $10^{-3}  $    & 2.28 \textcolor{red}{$ \times $} $10^{-3}  $ & 1.52 \textcolor{red}{$ \times $} $10^{-3}  $  &  1.09 \textcolor{red}{$ \times $} $10^{-3}  $  & 0.96 \textcolor{red}{$ \times $} $10^{-3}  $ &  0.88 \textcolor{red}{$ \times $} $10^{-3}  $   \\
                        \hline
                        3 & 3.19 \textcolor{red}{$ \times $}  $10^{-3}  $    & 1.89 \textcolor{red}{$ \times $} $10^{-3}  $ & 1.39 \textcolor{red}{$ \times $} $10^{-3}  $   &  0.9 \textcolor{red}{$ \times $} $10^{-3}  $  & 0.84 \textcolor{red}{$ \times $} $10^{-3}  $ &  0.79 \textcolor{red}{$ \times $} $10^{-3}  $   \\
                        \hline
                        5 & 2.89 \textcolor{red}{$ \times $} $10^{-3}  $    & 1.59 \textcolor{red}{$ \times $} $10^{-3}  $ & 1.29 \textcolor{red}{$ \times $} $10^{-3}  $  &  0.89 \textcolor{red}{$ \times $} $10^{-3}  $  & 0.71 \textcolor{red}{$ \times $} $10^{-3}  $ &  0.59 \textcolor{red}{$ \times $} $10^{-3}  $   \\
                        \hline
                        7 &  2.59 \textcolor{red}{$ \times $} $10^{-3}  $   & 1.54 \textcolor{red}{$ \times $} $10^{-3}  $ & 1.19 \textcolor{red}{$ \times $} $10^{-3}  $  &  0.79 \textcolor{red}{$ \times $} $10^{-3}  $  & 0.59 \textcolor{red}{$ \times $} $10^{-3}  $ &  0.39 \textcolor{red}{$ \times $} $10^{-3}  $   \\
                        \hline
                    \end{tabular}
        \caption{\textit{The critical intensity $ D_{c} $ of noise as a function of light intensity $ I_{0}$ and of stiffness $\mu$.}}
          \label{tab1}
        \end{center}
      \end{table}

Table \ref{tab1} collects the results on the critical value of the noise $ D_{c} $ as a function of the noise intensity  $ I_0 $ and $ \mu $.

 Evolution of the largest Lyapunov exponent
as a function of the noise for different values of the daylight duration $ D_{L} $ (i) of the external light and of the phase difference $ \phi $ (ii).
The duration of the daylight and the phase have no relevant
impact on the chaotic state in the wake/sleep system.
Thus the only parameters which affect the chaotic state
of our system are the stiffness and the intensity of the external stimulus.
Indeed we know that the condition to maintain the self-sustaining
oscillations in a van der Pol type oscillator is that the
dissipation coefficient (here, the stiffness $ \mu $) must be
relatively small \cite{tsaf}; which is in agreement with the results
 obtained since the increase in stiffness decreases the interval $] 0; D_{c}] $ system stability.

The critical value of the noise $ D_{c} $ indicates the maximum
of the noise for which the cycle is stable; Beyond this value
the disorder intervenes in the system and it becomes impossible
to predict not only the amplitude of the sleep (the desire to sleep)
but also its period. Since sleep plays a big role in the development
of the human body and many health problems are linked to a bad
wake sleep cycle \cite{lumiere}, noise could be one of the reasons for the human body to malfunction.
Quantitatively, this occurs that the critical illumination noise correlation $D_c$ is
pretty constant for a wide interval of the illumination intensity $I_0$, and
weakly decreases above a certain threshold, $I_0 \simeq 3$ for these parameters.

\section{Coherence resonance  }
\label{coherence}

In this Section it is highlighted the phenomenon of coherent
resonance in the sleep awakening system through numerical investigations.
Indeed, coherence resonance \cite{Nagumo, cham} entails that the stochastic
 system recovers some sort of regularity of the trajectories in spite of the fact that noise is increased.
In other words, one finds that the appropriated noise added to the
wake-sleep cycle can improve the system behavior.
How to measure these improvements calls for some measures of the coherence, as will be discussed below in Sect. \ref{tools} prior to the results on the coherence discussed in Sect. \ref{results}.

 \subsection{Tools to quantify coherent resonance}
 \label{tools}
 In order to study the effects of coherent resonance, the system is
 subjected to the sole influence of noise, Eq. (\ref{eq2}) with $ I_{0} = 0 $,
 that is to say in absence of excitatory signal. To characterize the coherent
 resonance effect, it is necessary to introduce a measuring tool such as
 the normalized autocorrelation $ C(\tau) $ \cite{Ctau1} defined by:
\begin{equation}
\label{eq4}
C (\tau) = \dfrac{\langle \tilde{x}(t) \tilde{x}(t-\tau)  \rangle}{\langle \tilde{x}(t)^{2}\rangle}, \qquad with \qquad \tilde{x}(t)= x(t)-\langle x(t)\rangle .
\end{equation}
Autocorrelation is calculated on the system variable $ x $ and
the variable $ \tilde{x} (t) $ corresponds to the oscillations
of the variable $ x (t) $ around its mean value $ \langle x (t) \rangle $.
Coherent resonance occurs when autocorrelation senses the most the noisy evolution.
In other words, one speaks of coherent (anti-coherent)
resonance when the amplitude of the autocorrelation decreases the least (more) quickly.
Coherent resonance can also be detected through the correlation time \cite{cham, Ctau1}:

\begin{equation}
\label{eq5}
\tau_{cor} = \frac{1}{Var(x(t))} \int_0 ^{+ \infty} C (\tau) d\tau, \qquad with \qquad Var(x(t)) = \langle x(t)^{2} \rangle - \langle x(t) \rangle^{2}.
\end{equation}
if this quantity exhibits a resonance for a particular value of the noise.

 \subsection{Results on coherence resonance }

\label{results}

Figures \ref{fig6} show, for the nonlinear damping  $ \mu = 0.23 $,
the autocorrelation for three values of the noise:  $ 8.5 $ \textcolor{red}{$ \times $}$ 10^{-5} $ (i), $ 9.9 $ \textcolor{red}{$ \times $}$ 10^{-5} $ (ii) and $ 15.5 $ \textcolor{red}{$ \times $}$ 10^{-5} $ (iii).
The autocorrelation amplitude decreases the fastest in Fig. \ref{fig4} (ii),
\emph{i.e.} for the intermediate value of noise  $ 9.9 $ \textcolor{red}{$ \times $}$ 10^{-5} $.
This determines that there is a value of the noise, close to  $ 9.9 $ \textcolor{red}{$ \times $}$ 10^{-5} $,
for which the noise depresses the most the autocorrelation of the wake-sleep system.
To support this result, we have plotted the evolution of the correlation
time $ \tau_{cor} $ for some values of $ \mu $ as a function
of the intensity of the noise in Fig. \ref{fig7}.
The following observations emerge:

 \begin{itemize}
 \item Assuming that the optimal desired effect of noise is to enhance the coherent resonance, or to achieve the highest correlation time, it is clear that the lowest noise level, ideally approaching zero, ($ D = 0 $) would be the most desirable.
 In fact, any increase of the noise initially reduces the correlation time, which deteriorates the coherence in the temporal evolution of the system.
 The interpretation in physiological consequences is not so obvious, it might be beneficial to shorten the correlation time, and thus to not ``propagate'' the disturbances in the subsequent cycles.
 In any case what we observe is that the system becomes more and more incoherent in the presence of noise
  \cite{cham,mbac},
  which creates the deterioration not only of the amplitude of the system (the desire to sleep) but also of the duration of sleep which can subsequently lead to sleep-related illnesses;
  \item There is an intensity of the noise for which the system is the least coherent corresponding to the correlation time is the lowest (Fig \ref{fig7}).
  In other words,  for an exact value of noise, the regularity of the response of the wake-sleep system is most degraded; indeed it is clear that in the system an anti-coherent resonance phenomenon has occurred, inasmuch there is an optimal noise above which noise degrades the regularity of the cycle.
 \item Finally we can see in this figure that the stiffness $ \mu $ of our system have a considerable impact on the phenomenon of coherent resonance. Indeed, whatever the value of the stiffness, the noise reflecting the anti-coherent resonance remains the same. Indeed, when $ \mu $ increases, the value of the noise reflecting the anti-coherent resonance decreases. Table \ref{tab2} gives us more information.
\end{itemize}

 \begin{table}[h!]
        \begin{center}
           \begin{tabular}[b]{|l|c|c|c|c|c|}
                        \hline
           Stiffness $ \mu$ & 0.023   &  0.13  & 0.23  & 0.39     & 0.46 \\
                         \hline
                Noise D &   $  19.9 $ \textcolor{red}{$ \times $}$ 10^{-5} $ & $  11.9 $ \textcolor{red}{$ \times $} $ 10^{-5} $ & $  9.9 $ \textcolor{red}{$ \times $}$ 10^{-5} $ & $  9.8 $ \textcolor{red}{$ \times $}$ 10^{-5} $ &  $  9.4 $ \textcolor{red}{$ \times $}$ 10^{-5} $ \\
                      \hline
                    \end{tabular}
        \caption{\textit{ Evolution of anti-coherence noise $D$ as function of the stiffness $ \mu$.}}
          \label{tab2}
        \end{center}
      \end{table}

Contrary to the phenomenon of coherent resonance in the system
of FithzHugh-Nagumo \cite{Nagumo} where there is a value of the
noise which improves the regularity of the system, there emerges
with regard to the wake-sleep cycle an optimal value of the noise
which deteriorates the regularity of the cycle. In other words,
noise contributes to the degradation of the evolution of the
 wake/sleep system in humans in the absence of daylight, as measured by the correlation time (\ref{eq5}).
This result is in agreement with those presented on the
van der Pol oscillators with limit multi-cycles \cite{cham};
the study of coherent resonance has been done on birhythmic systems
and it has been shown that there is a lower value of noise which
gives the best intensity necessary to obtain (anti) coherence resonance.

\section{Stochastic resonance }
\label{stochastic}
Stochastic resonance has been proposed \cite{Gammaitoni98} as the peculiar effect for which noise -- in the appropriate proportion, enhance the response of a nonlinear system to a periodic drive.
The concept of stochastic resonance has subsequently been expanded to describe any phenomenon where the presence of the right amount of noise in a nonlinear system is beneficial -- in a wide sense -- for the system response.
 The purpose of this Section is therefore first to briefly introduce some tools that quantify the ``benefits'', and then to apply the concepts to the wake-sleep system to ascertain whether noise can have some positive influence.

\subsection{Tools to quantify stochastic resonance}
 The word resonance in the term stochastic resonance was originally used because the change in output Signal-to-Noise Ratio (SNR) over noise intensity has a maximum (or minimum).
The SNR is introduced to quantify the contribution of the signal with respect to the noise; to do so one first defines the Root Mean Square (RMS) \cite{cham, Rms} by:

 \begin{equation}
\label{eq6}
RMS =  \int  x(t)x(t-t') dt',
\end{equation}
 which allows to determine the SNR \cite{cham, Rms1}
 \begin{equation}
\label{eq7}
SNR =  \dfrac{RMS(E\neq 0, D\neq 0)}{RMS(D\neq 0)},
\end{equation}
as the ratio of the RMS of noise and signal divided by the RMS in the presence of noise only.
This method was used in ref.\cite{cham} on Van Der Pol which was inspired by the different mathematical methods of the book \cite{Rms}.
When interpreted as an analysis of the sleep-awake cycle, one could surmise that it is a measure of the regularity of the sleep cycle \cite{Mendonca19}.
The noise value that maximizes the SNR is the one that improves the signal and creates resonance.
The simulations that follow aim to ascertain if noise can play such a constructive role.

 \subsection{Results of stochastic resonance}

 Figure \ref{fig8} displays the evolution of the SNR as
 a function of the noise intensity $ D $ for several values of the light intensity $ I_{0} $.
 There is a value of the noise, $ D = 3.4 $ \textcolor{red}{$ \times $}$ 10^{-5} $ for which
 the SNR is maximum, which highlights the phenomenon of stochastic resonance.
 When the light intensity $ I_{0} $ is increased, the curve is shifted upwards.
 Thus it is clear that the increase or decrease in the intensity of
 light applied to the system does not change the value of the
  noise which favors the wake-sleep cycle evolution in humans.
 To confirm this hypothesis, we plot in Fig. \ref{fig9} the SNR for different values of the duration of the light $ D_{L} $ (i) and of the phase $ \phi $ (ii). It appears that the duration and the phase of the light do not change the value of the noise for which there is resonance, namely  $ D = 3.4 $ \textcolor{red}{$ \times $}$ 10^{-5} $ . \\
 Therefore it appears that the parameters of the external light do not change the resonance peak which is in agreement with the results obtained on the van der Pol type oscillators \cite{mbac} where the noise value corresponding to the peak  resonance does not change when the correlation time is changed.
 Nevertheless let us note that the eigen parameters of the system can considerably change this resonance peak with regard to the bi-rhythmic oscillators \cite{cham}.

\section{Conclusions}
\label{conclusions}

Sleep is essential and helps maintain a healthy body. It is during sleep that the regeneration of the immune, nervous, skelatal and muscular systemtakes place \cite{lumiere}, restoring the reserve of adenosine triphosphate \cite{lit02} (ATP: molecule that acts as an energy current in the cell), and so forth; sleep must therefore follow a well-defined cycle.
The wake/sleep cycle studied here is modeled by a rhythmic van der Pol oscillator having higher order nonlinearity in the presence of an external excitation shaped like natural light.
As noise is inevitable in biological models, and illumination noise is nowadays a further source of disturbance, it is interesting to investigate the effects of an additive  noise \cite{blanc} on the circadian pacemaker.
The aim is to find out in which case noise might or might not promote the regularity of to sleep in this model.

To conduct this type of study, it is necessary to highlight several phenomena, namely: the bifurcation for the study of the presence of chaos, coherent and stochastic resonance for the beneficial effects of noise, both in the absence and in the presence of the external stimulus of light.
From the simulations it is possible to draw three main conclusions.

First, there is a critical noise $ D_c $ beyond which the system becomes chaotic; so above this noise level an utter disorder occurs in the wake/sleep cycle and it becomes impossible to predict the amplitude and period of sleep.
This critical noise varies according to the stiffness $ \mu $ of the system and the light intensity $ I_{0} $ but remains constant for the other parameters ($ D_{L} $, $ \phi $) of the external stimulus.

The measurement of the correlation rate in the absence of the external stimulus makes it possible to observe that there is an appropriate dose of the noise for which the regularity of the response of the wake-sleep system is the most degraded; hence the demonstration of the phenomenon of anti-coherent resonance.
Indeed, there exists a value of the noise  $ D$ depending of the value of stiffness $ \mu $ for which  the regularity of the cycle is rather degraded. Table \ref{tab2}  gives us more information. \\
Finally, about stochastic resonance, the evaluation of the SNR allows to determine that there is a value of the noise, $ D = 3.4 $ \textcolor{red}{$ \times $} $ 10^{-5} $, which favors the evolution of our system.
This implies that noise can be beneficial for the wake-sleep cycle model in the presence of natural light.
In fact, this noise value does not vary according to the stiffness and the parameters of natural light.

In short, in this van der Pol model noise can cause a dysfunction in the wake/sleep cycle in humans by making the cycle completely unpredictable (amplitude and period of the cycle), which can lead to certain sleep-related diseases over time.
Nevertheless there is an optimal noise which can improve the evolution in the presence of the illumination stimulus.
We hope that this work might prove of importance in biology, for it allows us to better understand the behavior of circadian rhythms and gives results that can be exploited to model the sleep/awake cycle in different illumination conditions, including the effects of artificial light oscillations.

\section*{Conflict of Interest}
The authors declare that they have no conflict of interest.


\newpage

\newpage

\begin{figure}
\begin{center}
\includegraphics[height=7.0cm,width=14.0cm]{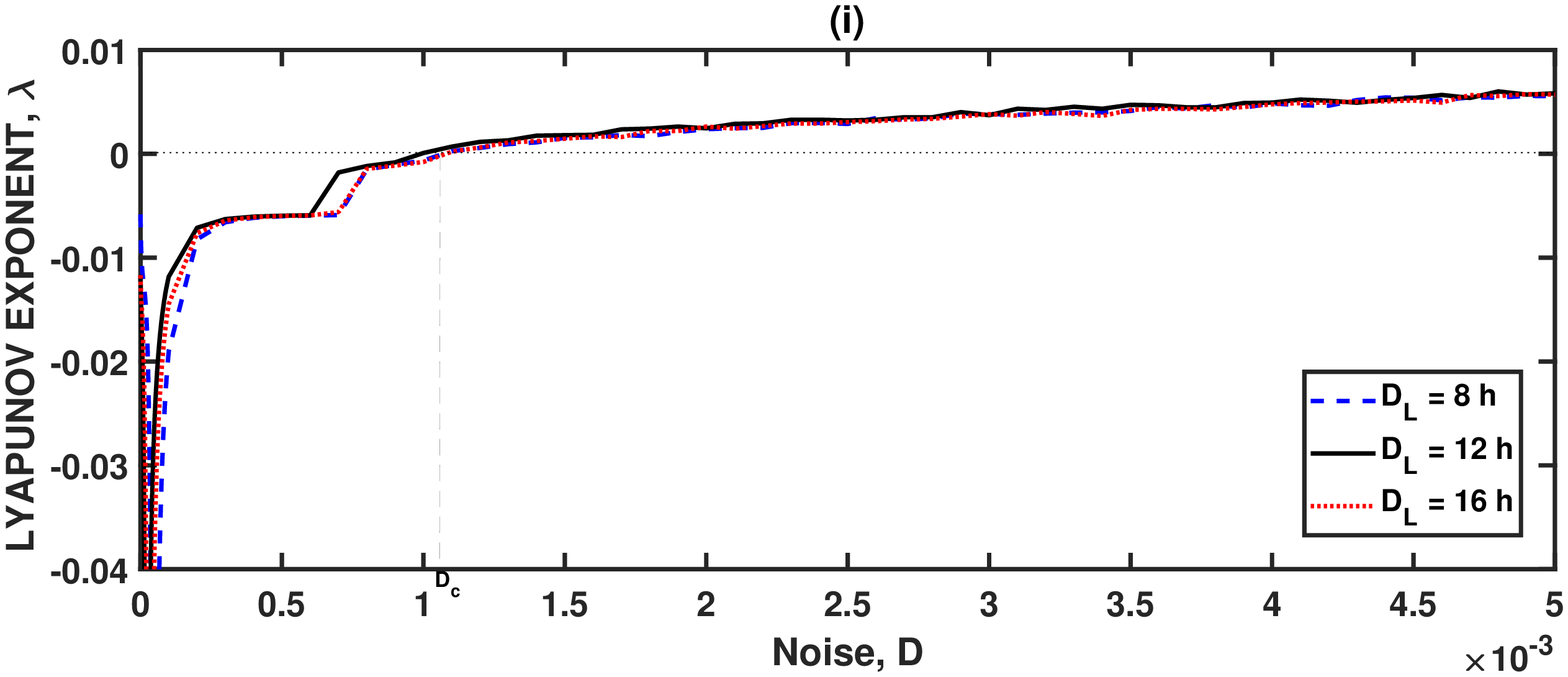}
\includegraphics[height=7.0cm,width=14.0cm]{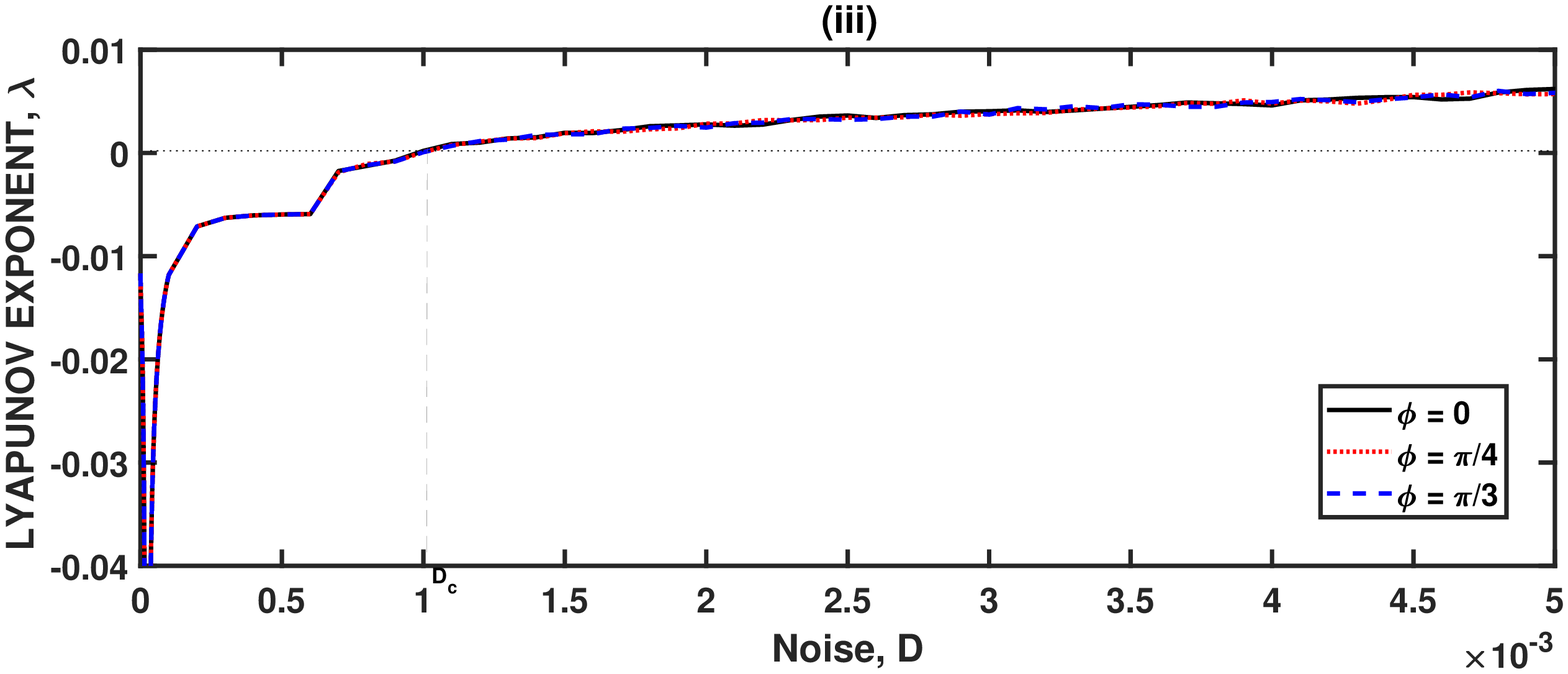}
\caption{\it  (i) Effects of the duration of the daylight $D_{L}$ on
 the variation of the largest Lyapunov exponent  versus  the noise intensity $D$
  with $ \phi = 0 $. (ii) Effects of $\phi$ on the variation of the largest Lyapunov exponent
  versus the noise intensity $D$.
The other parameters used are : $ I_{0} = 2.5 $, $\tau_{x} = 24.2$ . }
\label{fig5}
\end{center}
\end{figure}

\begin{figure}
\begin{center}
\includegraphics[height=7.0cm,width=14.0cm]{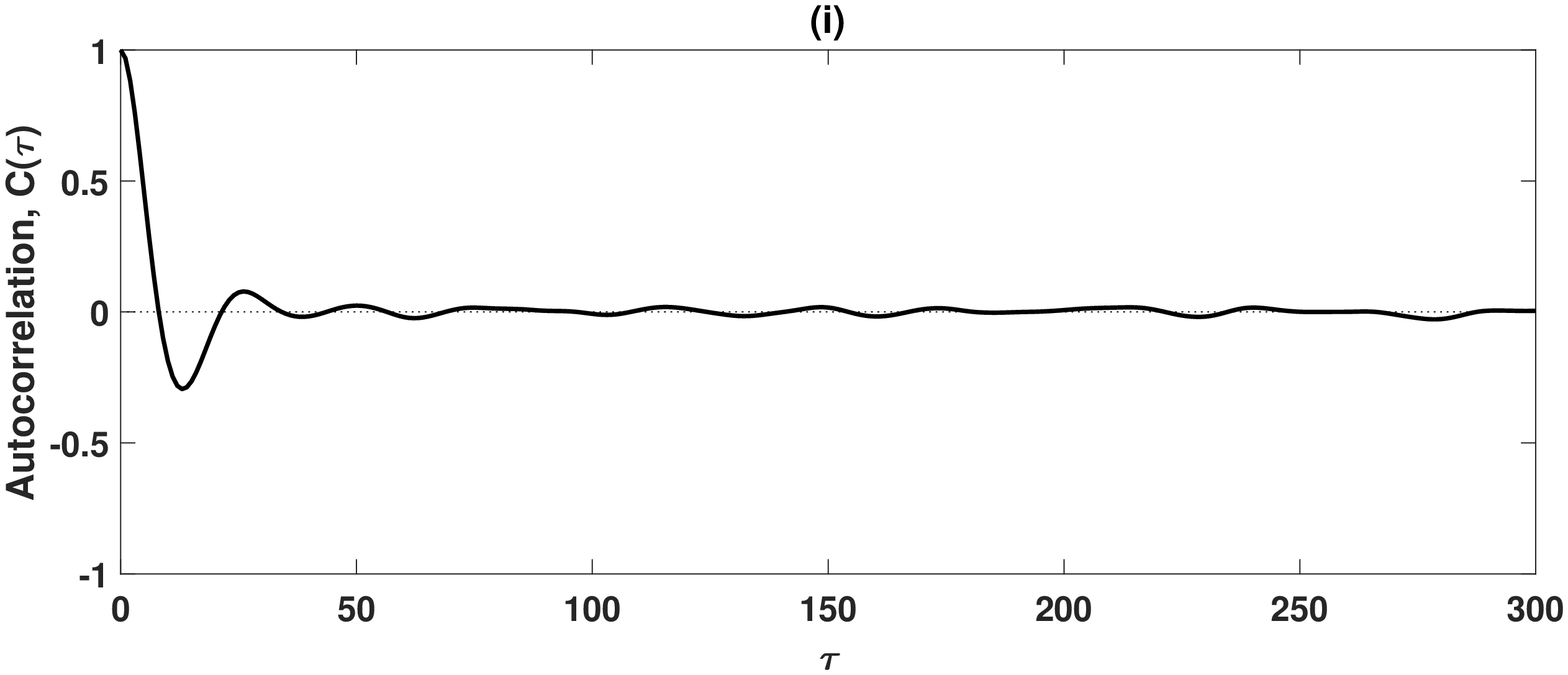}
\includegraphics[height=7.0cm,width=14.0cm]{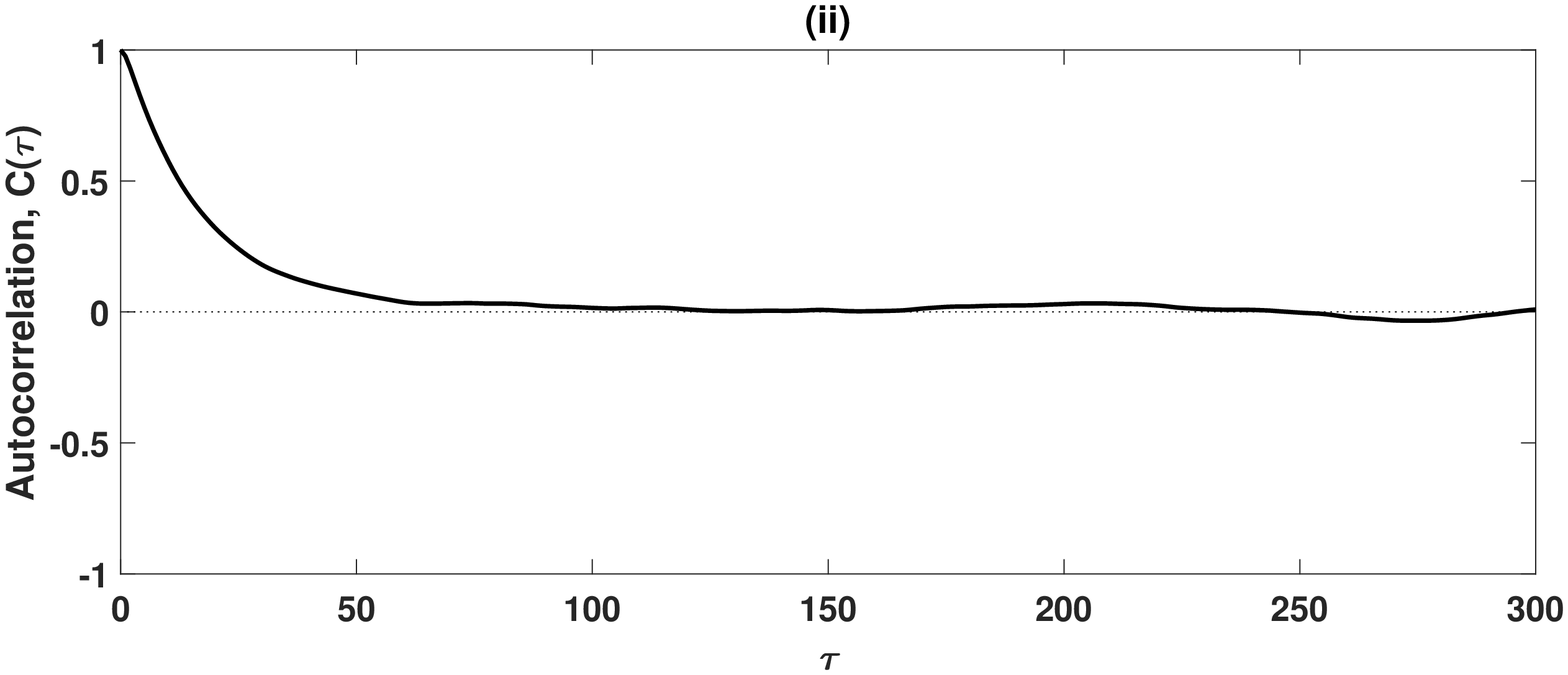}
\includegraphics[height=7.0cm,width=14.0cm]{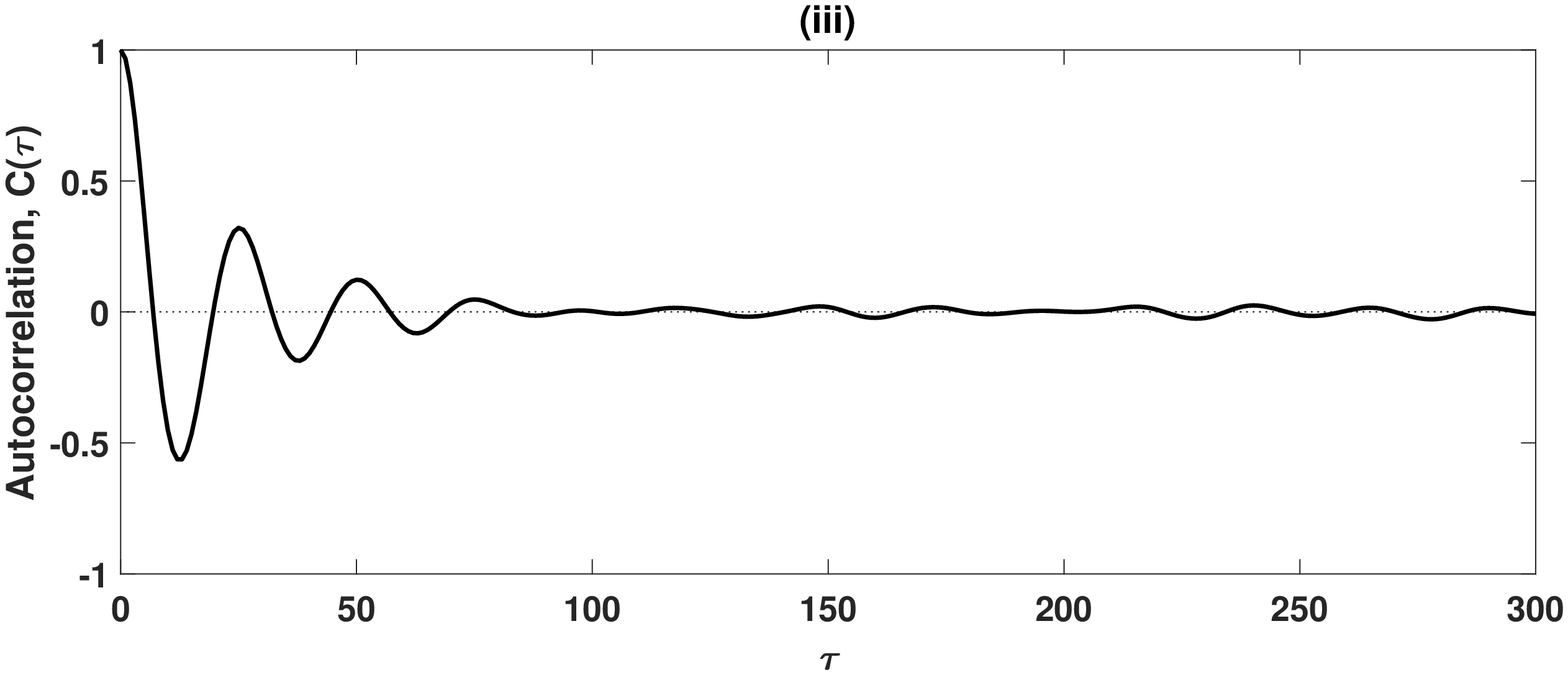}
\caption{\it Effects of the noise intensity $D$ of the
variation of the autocorrelation function $C$ versus $\tau$; (i): $D = 8.5  $ \textcolor{red}{$ \times $}$ 10^{-5}$;
(ii): $ D= 9.9  $ \textcolor{red}{$ \times $}$ 10^{-5} $ and (iii): $ D= 15.5  $ \textcolor{red}{$ \times $}$ 10^{-5} $.
The other parameters used are: $ \mu = 0.23 $, $ I_{0} = 0 $, $\tau_{x} = 24.2$. }
\label{fig6}
\end{center}
\end{figure}

\begin{figure}
\begin{center}
\includegraphics[height=6.0cm,width=14.0cm]{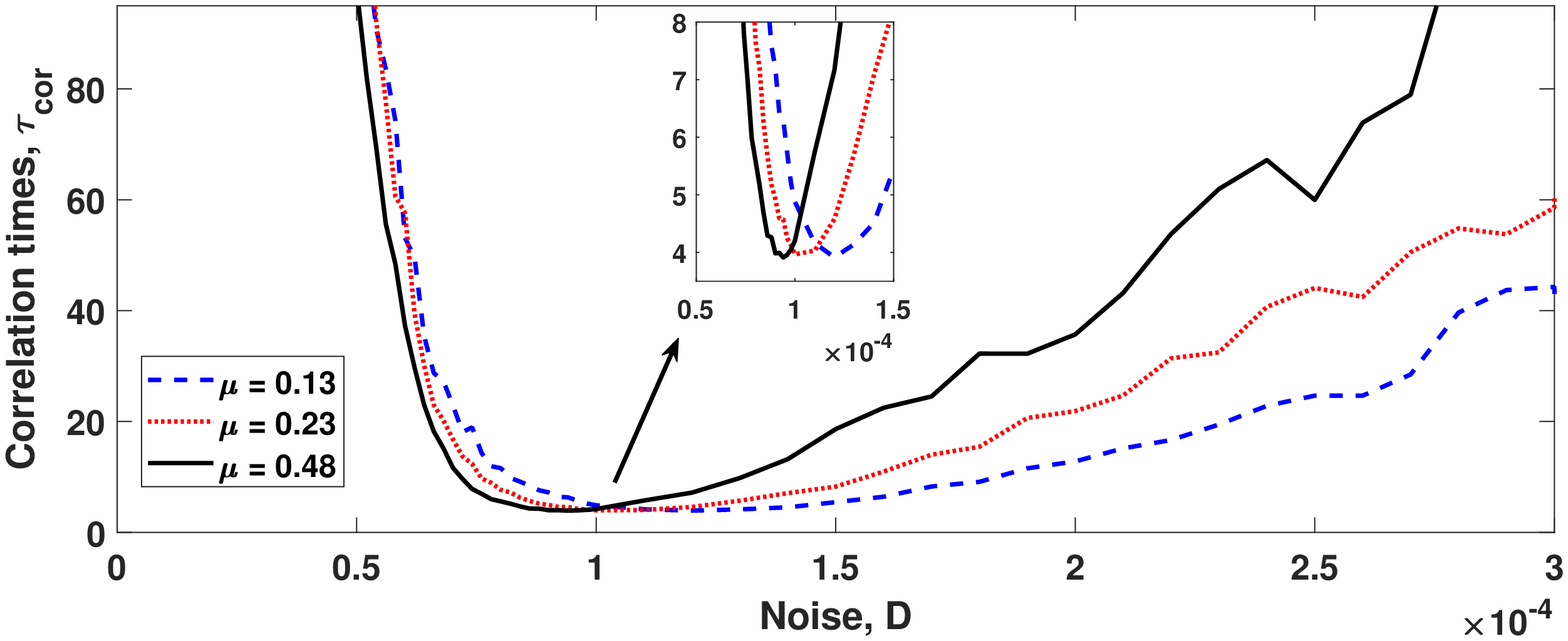}
\caption{\it  Effects of the stiffness $\mu$ on the variation of
the correlation time $\tau_{cor}$ versus the noise intensity  $D$.
The other parameters used are defined in Fig. \ref{fig6}.}
\label{fig7}
\end{center}
\end{figure}

\begin{figure}
\begin{center}
\includegraphics[height=7.0cm,width=14.0cm]{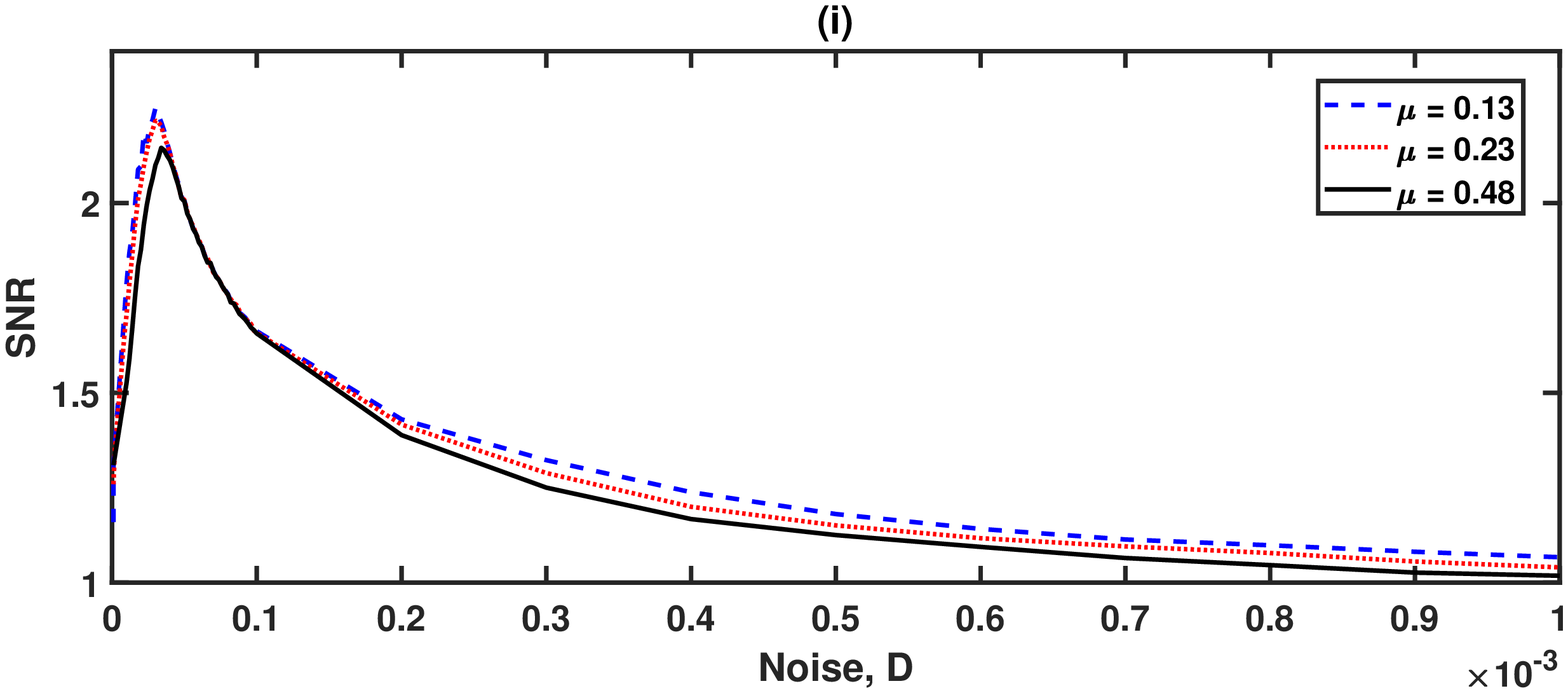}
\includegraphics[height=7.0cm,width=14.0cm]{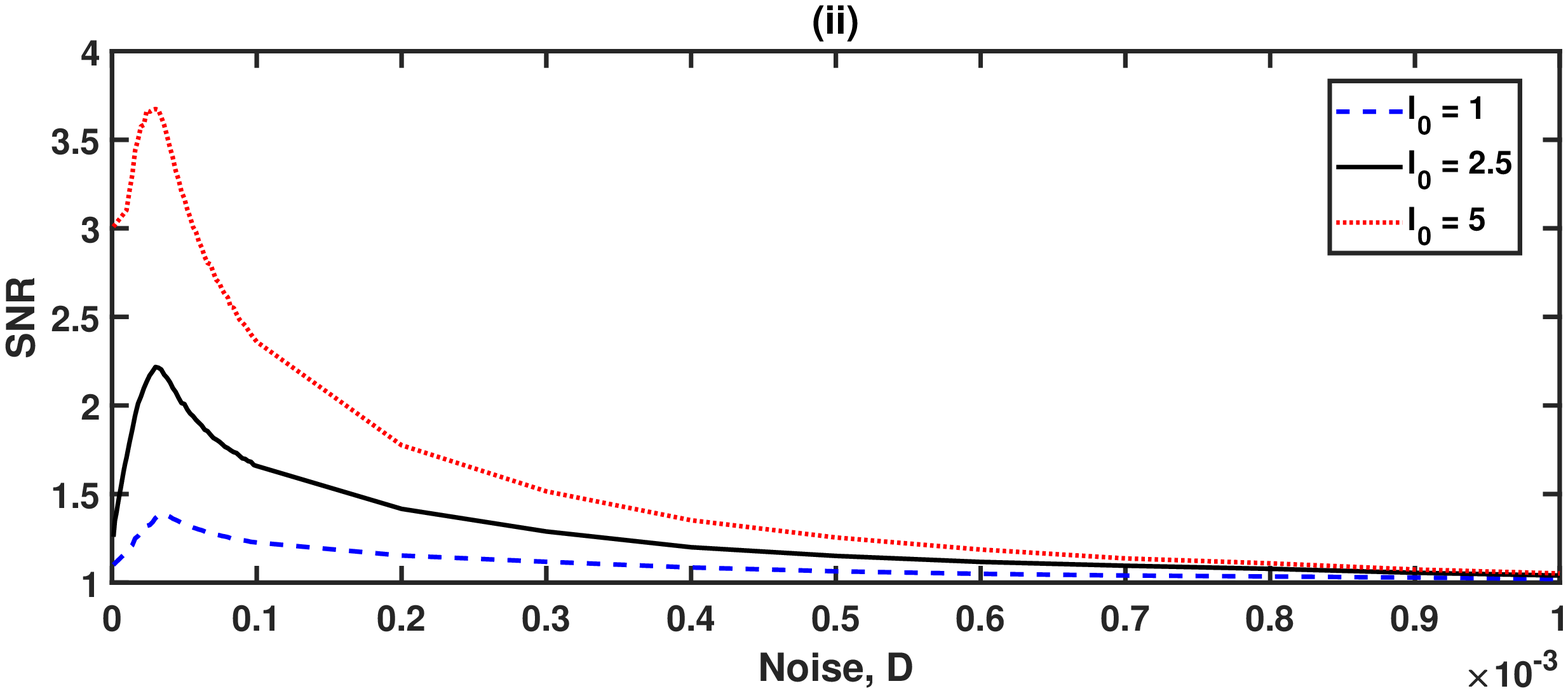}
\caption{\it (i) Effects of the stiffness $\mu$ on the variation
of the signal to noise ratio $SNR$ versus  the noise intensity
$D$.  (ii) Effects of the light intensity $I_{0}$ on the variation
of the signal to noise ratio $SNR$ versus  the noise intensity
$D$.
The other parameters used are: $\tau_{x} = 24.2$ ,$ D_{L} = 12 h $ and $ \phi = 0 $.}
\label{fig8}
\end{center}
\end{figure}

\begin{figure}
\begin{center}
\includegraphics[height=7.0cm,width=14.0cm]{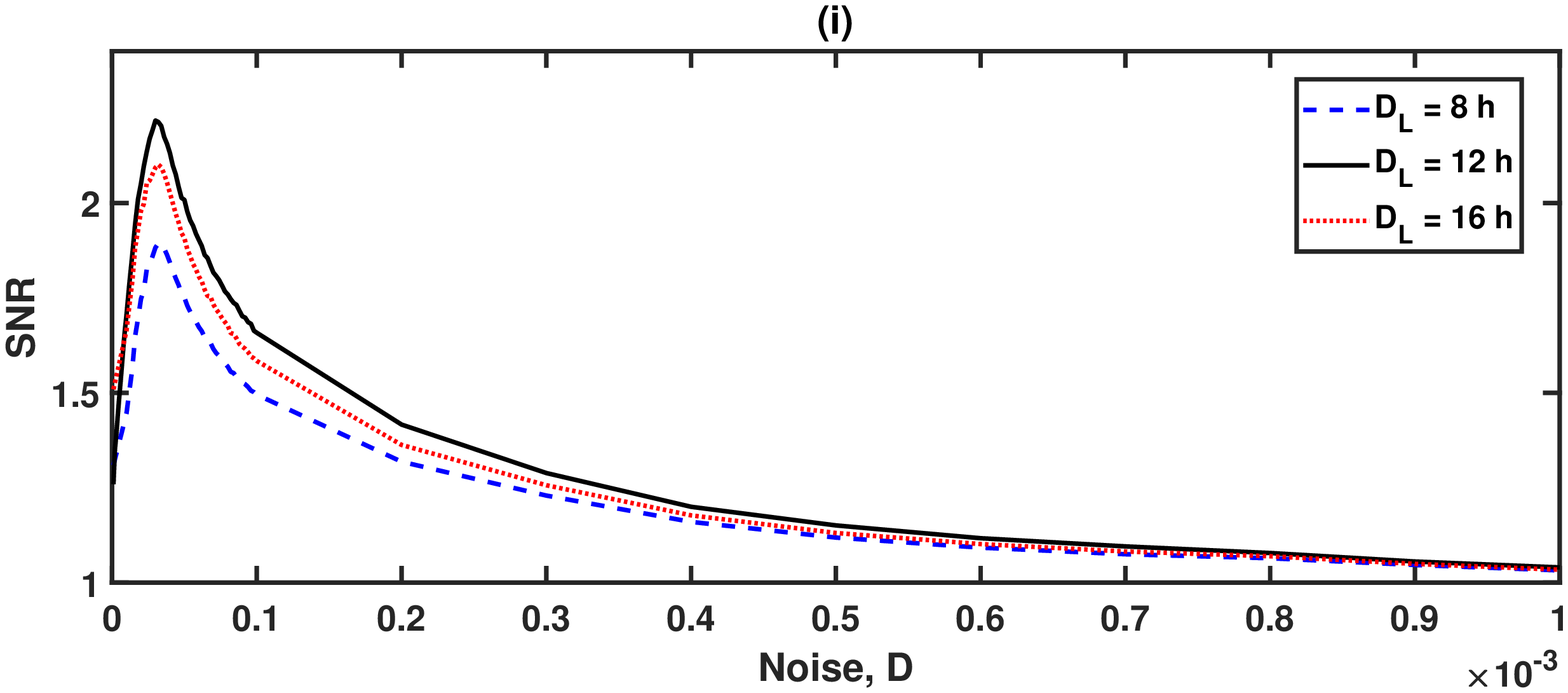}
\includegraphics[height=7.0cm,width=14.0cm]{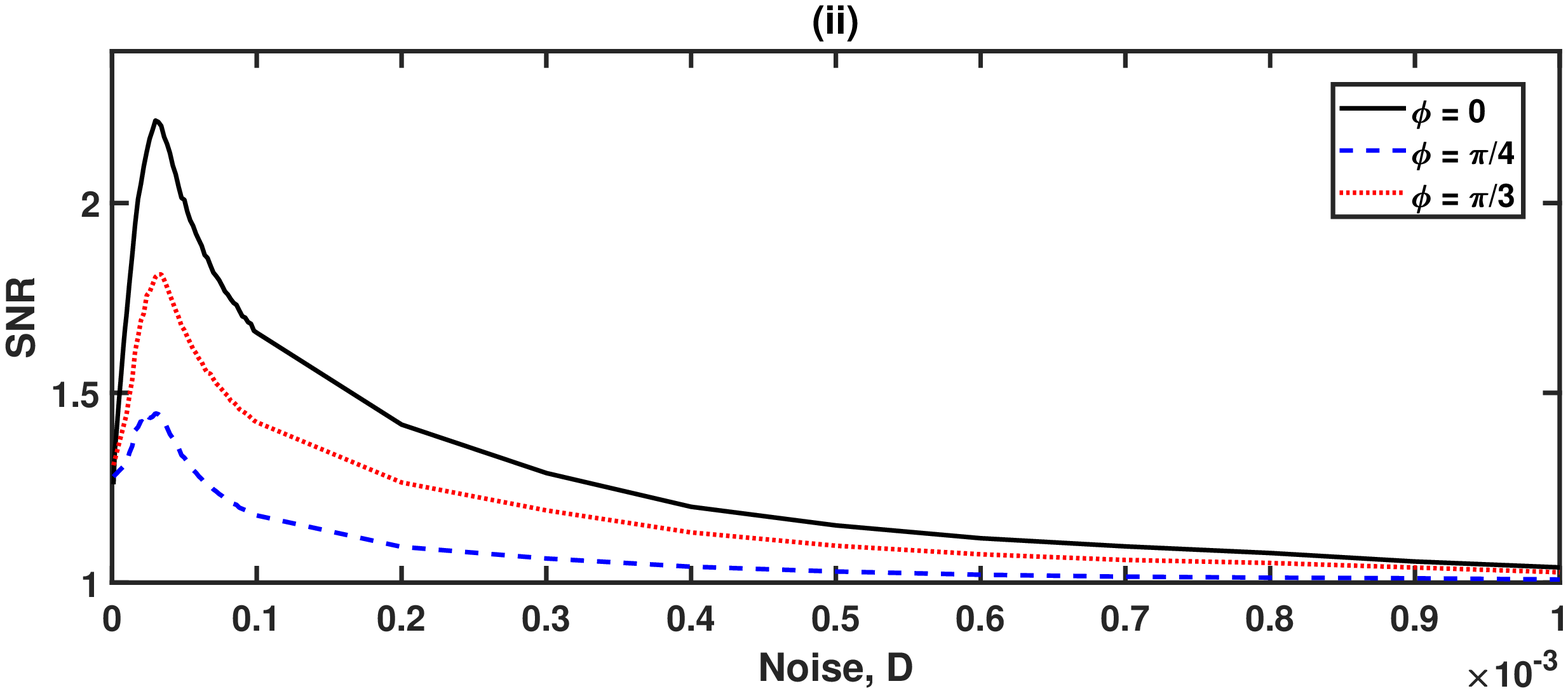}
\caption{\it (i) Effects of the daylight duration $D_L$ on the variation of
 the signal to noise ratio SNR versus the noise intensity $D$ for $ I_{0} = 2.5 $,$ \phi = 0 $).
  (ii) Effects of $\phi$  on the variation of
 the signal to noise ratio SNR versus the noise intensity $D$ for  $ I_{0} = 2.5 $, $ D_{L} = 12 h $.
The other parameters used are defined in Fig. \ref{fig1}.}
\label{fig9}
\end{center}
\end{figure}

\end{document}